\documentclass[10pt,aps,twocolumn,prb,superscriptaddress,longbibliography,floatfix]{revtex4-2}

\usepackage{titlesec}
\usepackage[T1]{fontenc}
\titleformat{\section}
  {\centering\normalfont\bfseries\MakeUppercase} % Normal font size
  {\thesection}{1em}{}                           % Numbering format

\usepackage{amsmath,amsfonts,amssymb}
\usepackage{graphicx,color}
\usepackage{hyperref}
\usepackage{epstopdf}

\usepackage{multirow}
\usepackage{hhline}
\usepackage{ulem}
\usepackage{mathtools}
\usepackage[caption=false]{subfig}
\usepackage{soul}
\usepackage{booktabs}
\hypersetup{hidelinks,colorlinks=true,allcolors=BLUE}
\usepackage{hyperref}  % Enables hyperlinks in the document

\usepackage{doi}       % Handles DOI links properly

\makeatletter

\@ifundefined{textcolor}{}
{%
 \definecolor{BLACK}{gray}{0}
 \definecolor{WHITE}{gray}{1}
 \definecolor{RED}{rgb}{1,0,0}
 \definecolor{GREEN}{rgb}{0,1,0}
 \definecolor{BLUE}{rgb}{0,0,1}
 \definecolor{CYAN}{cmyk}{1,0,0,0}
 \definecolor{MAGENTA}{cmyk}{0,1,0,0}
 \definecolor{YELLOW}{cmyk}{0,0,1,0}
}

\makeatother
\usepackage{float}

\begin{document}

\title{Demonstrating magnetic field robustness and reducing temporal $T_1$ noise in transmon qubits through magnetic field engineering}

\author{Bektur Abdisatarov}
\email{Corresponding author: bektur@fnal.gov}
\affiliation{Superconducting Quantum Materials and Systems Center,
Fermi National Accelerator Laboratory (FNAL), Batavia, IL 60510, USA}
\affiliation{Department of Electrical and Computer Engineering, Old Dominion University, Norfolk, Virginia 23529, USA and Applied Research Center, 12050 Jefferson Avenue, Newport News, Virginia 23606, USA}
\author{Tanay Roy}
\email{Corresponding author: roytanay@fnal.gov}
\affiliation{Superconducting Quantum Materials and Systems Center,
Fermi National Accelerator Laboratory (FNAL), Batavia, IL 60510, USA}
\author{Daniel Bafia}
\affiliation{Superconducting Quantum Materials and Systems Center,
Fermi National Accelerator Laboratory (FNAL), Batavia, IL 60510, USA}
\author{Roman Pilipenko}
\affiliation{Superconducting Quantum Materials and Systems Center, Fermi National Accelerator Laboratory (FNAL), Batavia, IL 60510, USA}
\author{Matthew Julian Dubiel}
\affiliation{Superconducting Quantum Materials and Systems Center, Fermi National Accelerator Laboratory (FNAL), Batavia, IL 60510, USA}
\author{David van Zanten}
\affiliation{Superconducting Quantum Materials and Systems Center, Fermi National Accelerator Laboratory (FNAL), Batavia, IL 60510, USA}
\author{Shaojiang Zhu}
\affiliation{Superconducting Quantum Materials and Systems Center, Fermi National Accelerator Laboratory (FNAL), Batavia, IL 60510, USA}
\author{Mustafa Bal}
\affiliation{Superconducting Quantum Materials and Systems Center, Fermi National Accelerator Laboratory (FNAL), Batavia, IL 60510, USA}
\author{Grigory Eremeev}
\affiliation{Superconducting Quantum Materials and Systems Center, Fermi National Accelerator Laboratory (FNAL), Batavia, IL 60510, USA}
\author{Hani Elsayed-Ali}
\affiliation{Department of Electrical and Computer Engineering, Old Dominion University, Norfolk, Virginia 23529, USA and Applied Research Center, 12050 Jefferson Avenue, Newport News, Virginia 23606, USA}
\author{Akshay Murty}
\affiliation{Superconducting Quantum Materials and Systems Center,
Fermi National Accelerator Laboratory (FNAL), Batavia, IL 60510, USA}
\author{Alexander Romanenko}
\affiliation{Superconducting Quantum Materials and Systems Center, Fermi National Accelerator Laboratory (FNAL), Batavia, IL 60510, USA}
\author{Anna Grassellino}
\affiliation{Superconducting Quantum Materials and Systems Center, Fermi National Accelerator Laboratory (FNAL), Batavia, IL 60510, USA}

\date{\today}

\begin{abstract}

The coherence of superconducting transmon qubits is often disrupted by fluctuations in the energy relaxation time ($T_1$), limiting their performance for quantum computing. While background magnetic fields can be harmful to superconducting devices, we demonstrate that both trapped magnetic flux and externally applied static magnetic fields can suppress temporal fluctuations in $T_1$ without significantly degrading its average value or qubit frequency. Using a three-axis Helmholtz coil system, we applied calibrated magnetic fields perpendicular to the qubit plane during cooldown and operation. Remarkably, transmon qubits based on tantalum-capped niobium (Nb/Ta) capacitive pads and aluminum-based Josephson junctions (JJs) maintained $T_1$ lifetimes near 300~$\mu$s even when cooled in fields as high as 600~mG. Both trapped flux up to 600~mG and applied fields up to 400~mG reduced $T_1$ fluctuations by more than a factor of two, while higher field strengths caused rapid coherence degradation. We attribute this stabilization to the polarization of paramagnetic impurities, the role of trapped flux as a sink for non-equilibrium quasiparticles (QPs), and partial saturation of fluctuating two-level systems (TLSs). These findings challenge the conventional view that magnetic fields are inherently detrimental and introduce a strategy for mitigating noise in superconducting qubits, offering a practical path toward more stable and scalable quantum systems.

\end{abstract}

\maketitle

\section{Introduction}
Fluctuations in $T_1$ of superconducting transmon qubits can disrupt calibration protocols and degrade gate fidelity in quantum processors~\cite{krantz2019quantum, kjaergaard2020superconducting, siddiqi2021engineering, mcdermott2009materials, burnett2019decoherence}. These fluctuations arise from a complex interplay of microscopic noise sources~\cite{klimov2018fluctuations, bejanin2021interacting, yoo2023fluctuations}, including TLSs in amorphous dielectrics~\cite{mueller2015interacting, abdisatarov2024direct, bafia2023oxygen, oh2024structure}, non-equilibrium QPs~\cite{catelani2012decoherence, aumentado2023quasiparticle}, and low-frequency magnetic flux noise associated with spin impurities and surface defects~\cite{kumar2016origin, szczesniak2021magnetic, rower2023evolution}. As coherence times continue to improve, identifying and mitigating these stochastic processes become increasingly critical for the scalability of quantum hardware. In particular, recent studies show that $T_1$ fluctuations often scale with the mean value $T_1$, highlighting the limitations of short-term measurements in evaluating the effects of fabrication or material improvements~\cite{kjaergaard2020superconducting, siddiqi2021engineering, klimov2018fluctuations}. Accurate characterization of these noise sources requires long-term statistical sampling due to their nonstationary and correlated nature.

The relaxation time \( T_1 \) in superconducting transmon qubits based on Nb/Ta films is predominantly limited by dielectric losses, particularly those arising from TLS associated with surface oxides and lossy dielectric substrates~\cite{siddiqi2021engineering, bal2024systematic}. Native oxides such as Nb\(_2\)O\(_5\) and Ta\(_2\)O\(_5\) are known to host TLSs and are considered among primary sources of limitations of \( T_1 \) performance~\cite{bal2024systematic}.

However, TLS-related losses are not the sole contributors to decoherence. Conductive losses arising from surface resistance (\( R_s \)) also contribute to energy dissipation. For niobium films at millikelvin temperatures, surface resistance has been measured to be \( R_s \approx 4.2\,\mathrm{n}\Omega \)~\cite{abdisatarov2024direct}. In addition to dielectric and conductive losses, further dissipation can originate from magnetic flux trapped in the superconducting film. Although the Meissner effect ideally expels magnetic fields during the superconducting transition, real materials inevitably contain pinning centers, such as defects or structural inhomogeneities, that enable partial flux penetration and the formation of vortices.
These trapped magnetic vortices can interact with the qubit’s microwave field and introduce extra losses~\cite{aull2012trapped, PhysRevAccelBeams.22.022001}.
Thus, total surface resistance can be expressed as \( R_s + R_{\mathrm{tf}} \), where \( R_{\mathrm{tf}} \) accounts for the loss associated with trapped flux. The number of trapped vortices is approximately given by

\[
N = \frac{B \times A}{\Phi_0}, \quad \Phi_0 \approx 2.07 \times 10^{-15}~\mathrm{Wb},
\]

\noindent
where \( B \) is the magnetic field applied during the cooldown, \( A \) is the surface area of the superconducting structure, and \( \Phi_0 \) is the magnetic flux quantum. As the vortex density increases, so does \( R_{\mathrm{tf}} \), leading to a reduction in \( T_1 \).

This phenomenon has been extensively studied in superconducting radio-frequency (SRF) cavities made of bulk niobium operating at 1.5--2~K~\cite{romanenko2014ultra,posen2016efficient} and, more recently, our group extended these studies to the quantum regime at millikelvin temperatures~\cite{bafia2025quantifying}. We find that levels of trapped magnetic flux above \( \sim 100\,\mathrm{mG} \) can reduce the quality factor of the SRF cavity and start to matter as compared to the dominant oxide losses, with each milligauss of trapped flux during cooldown contributing approximately \( 2\,\mathrm{n}\Omega \) of additional surface resistance~\cite{bafia2025quantifying}.

While these results quantify the impact of magnetic flux in SRF cavities, the role of magnetic fields in superconducting qubit systems is more complex. SRF cavities are a simple system composed of bulk niobium and niobium oxide on the surface, while transmon qubits contain multiple interfaces, junction and substrate materials, and the different geometries and participation ratios make qubits less sensitive than cavities to trapped magnetic field losses. With this study we aim at quantifying the impact of magnetic fields on high coherence transmon qubits performance. Although much attention has been focused on dielectric and quasiparticle-related losses, the effects of magnetic fields on paramagnetic impurities remain less explored yet a potentially influential decoherence channel~\cite{lee2014identification,sendelbach2008magnetism,cava1991electrical,pritchard2024suppressed}. Flux vortices in superconducting films can introduce dissipation and frequency instability, yet recent studies have uncovered interesting effects: weak magnetic fields can, in some cases, enhance coherence in low-$T_1$ qubits. Such effects, observed in aluminum qubit devices~\cite{wang2014measurement} and titanium nitride qubit devices~\cite{schneider2019transmon}, remain difficult to interpret due to the limited resolution afforded by the short coherence times (\( T_1 \)$<30~\mu$s).

High-coherence ($T_1 > 100~\mu$s) transmon qubits constructed at the SQMS Center from Nb/Ta capacitor pads and aluminum JJs now offer a new regime for testing decoherence with high sensitivity~\cite{bal2024systematic}. In this work, we uncover novel findings: both trapped magnetic flux and externally applied static fields can suppress temporal fluctuations in $T_1$ without degrading $T_1$'s average value, provided that field amplitudes remain below critical thresholds.

Using a precision 3-axis Helmholtz coil system, we systematically applied perpendicular magnetic fields during cooldown to trap flux in the pads, as well as during $T_1$ measurements. We found that trapping flux up to 600\,mG and applying static magnetic fields up to 400\,mG stabilized $T_1$ over timescales of several hours, without affecting the mean $T_1$ value. We attribute this stabilization to a combination of paramagnetic impurity polarization, the role of trapped flux as a sink for non-equilibrium QPs, and partial saturation of TLSs. These results reveal a mechanism by which magnetic fields, traditionally regarded detrimental to superconducting coherence, can instead be harnessed to suppress quantum noise. Our findings redefine the role of magnetic environments in qubit operation and point toward new strategies for stabilizing superconducting quantum hardware.

\maketitle
\section{Device description and Experimental setup}

\begin{figure}[h]
\centering\includegraphics[width=\columnwidth]{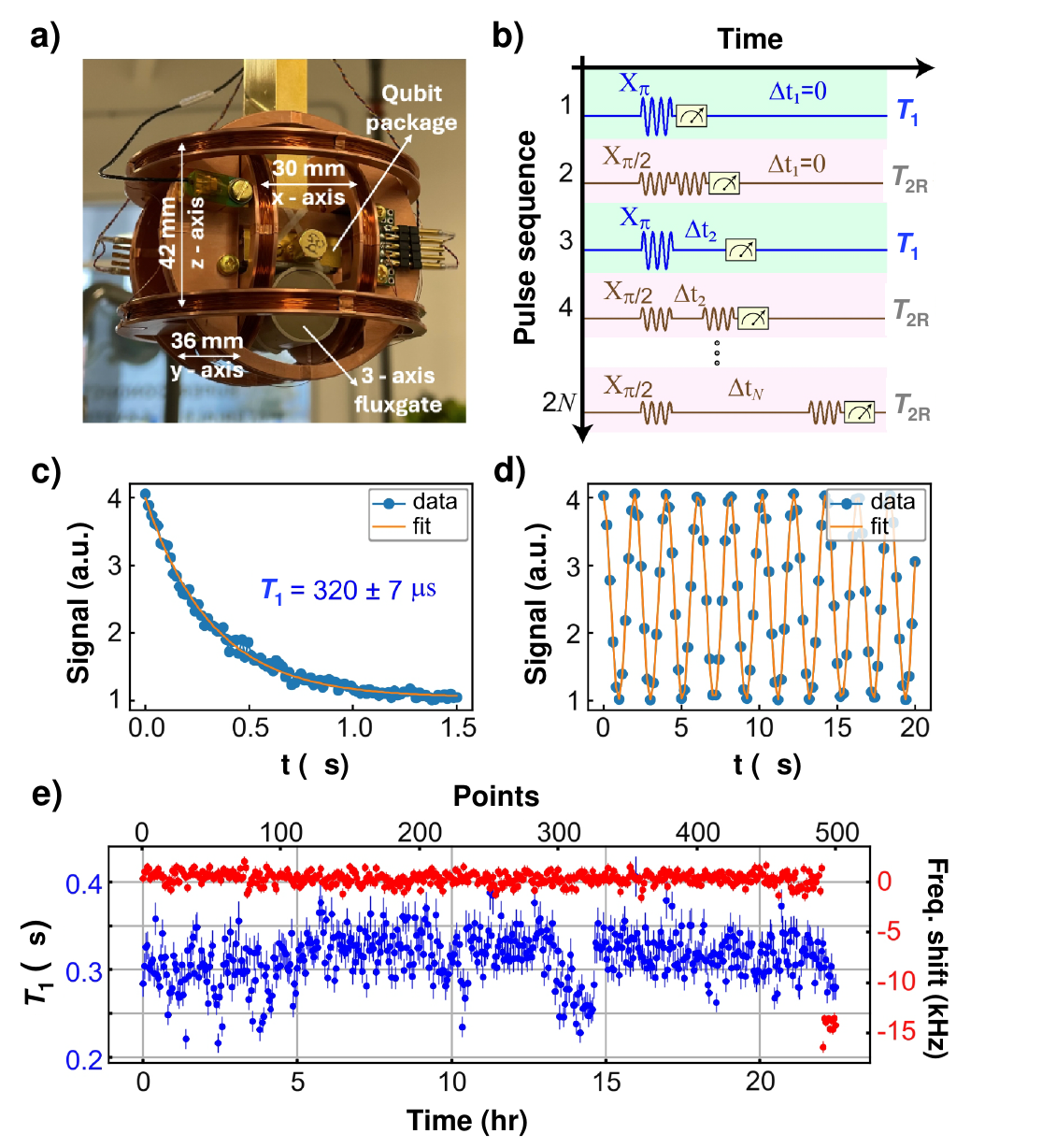}
\caption{Helmholtz coil setup and qubit measurement procedures. a) Three orthogonal Helmholtz coils independently apply magnetic fields along the \( x \), \( y \), and \( z \) -axes, with the qubit package oriented in the \( x \)-\( y \) plane. A 3-axis fluxgate sensor monitors the local magnetic field near the qubit chip. b) Pulse sequence for interleaved \( T_1 \) and Ramsey measurements, with each point averaged over 500 repetitions. Example traces of c) \( T_1 \), fit to an exponential, and d) Ramsey fringes, fit to an exponentially decaying sine function, are shown. e) The sequence in b) is repeated 500 times to track temporal fluctuations in \( T_1 \), qubit frequency, and fitting errors.}
\label{fig:Device}
\end{figure}

To generate a controlled magnetic environment at millikelvin temperatures, we designed and implemented a three-axis Helmholtz coil system inside a BlueFors XLD-type commercial dilution refrigerator (DR). Each axis of the system consists of a pair of circular copper structures, around which a superconducting Nb wire coated with copper was wound, as illustrated in FIG.~\ref{fig:Device}a. Each pair comprises of 100 turns (50 turns each) providing symmetrical field generation. Copper was selected for its high thermal conductivity, ensuring effective heat sinking, the use of superconducting Nb wire minimizes resistive heating during magnetic field application.

The radii of the coils were optimized for field uniformity at the qubit location, with values of 30\,mm, 36\,mm, and 42\,mm for the \( x \), \( y \), and \( z \)-axes, respectively. The qubit chip box was positioned at the geometric center of the coil assembly to ensure magnetic field uniformity. A three-axis fluxgate magnetometer was mounted directly above the qubit chip box to enable precise, \textit{in-situ} measurements of the applied magnetic field. Field homogeneity was verified using COMSOL Multiphysics simulations.

To study the relaxation dynamics of qubits and their fluctuations, we selected two Nb capped with Ta transmon qubit chip sets previously characterized in the literature~\cite{bal2024systematic}. All chips were fabricated on annealed HEMEX-grade sapphire substrates and included aluminum-based JJs. The coherence measurements were performed using concurrent $T_1$ and Ramsey experiments where the individual points in the time traces for the curves were interleaved.  The pulse sequence is shown in FIG.~\ref{fig:Device}b. $T_1$ data was fitted to the expression $A \exp{-t/T_1} + B$ to extract relaxation times. Similarly, Ramsey fringe data was fitted to $A \exp{-t/T_{2R}} \sin(\omega_R t + \phi) + B$ to extract the qubit frequency and dephasing parameters. Example traces of $T_1$ and Ramsey fringes are shown in FIG.~\ref{fig:Device}c and ~\ref{fig:Device}d. For statistical analysis, the mean (M) and mean absolute deviation (MAD) were used to characterize the qubit's $T_1$ and its temporal fluctuations.

The first chip set, Qubit Chip Box 1 ($QCB_1$), comprised of eight transmon chips with Nb films deposited at room temperature (RT) and capped with Ta. We selected two qubits from this set: qubit 1 ($q_1$) and qubit 2 ($q_2$), with capacitor pad dimensions of $120 \times 510\,\mu$m and $170 \times 890\,\mu$m, and pad spacings of $20\,\mu$m and $180\,\mu$m, respectively.

The second chip set, Qubit Chip Box 2 ($QCB_2$), contained eight transmon chips with Nb films deposited at 800\,°C and similarly capped with Ta. From this set, we selected qubit 3 ($q_3$), with $120 \times 510\,\mu$m capacitor pads and $20\,\mu$m pad spacing. Measurements of $T_1$ for $q_1$, $q_2$, and $q_3$ were performed at 8\,mK for durations of 12, 24, and 15 hours, respectively, and were consistent with previously reported results~\cite{bal2024systematic}.

To examine the influence of trapped magnetic flux on qubit relaxation, we performed a series of controlled cooldowns for $q_1$ and $q_2$. Static magnetic fields of 400\,mG, 600\,mG, 800\,mG, and 1000\,mG were applied along the z-axis during cooldown from 15\,K to 3\,K. The field was then turned off and the DR was further cooled to 8\,mK. $T_1$ was measured for 12 and 24 hours for $q_1$ and $q_2$, respectively.

To evaluate the thermal effects on the trapped flux, a field of 800\,mG was applied during the cooldown from 15\,K to 3\,K then turned off. The qubits were further cooled to the base temperature (8\,mK), followed by a temperature sweep up to 200\,mK and back down to 8\,mK. This allowed us to observe the evolution of $T_1$ in the presence of pre-trapped flux across temperature cycles.

Finally, to study the impact of actively applied magnetic fields at base temperature, we applied static magnetic fields of 100\,mG, 200\,mG, and 400\,mG along the z-axis during $T_1$ measurements of $q_3$. This comprehensive experimental approach enabled a detailed investigation of both trapped and externally applied magnetic field effects on transmon qubit coherence.

\section{Results}

We systematically studied the effects of both trapped and actively applied magnetic fields on \( T_1 \) and its temporal fluctuations in three transmon qubits. The temporal stability of \( T_1 \) was quantified using M and MAD in long-duration measurements. No correlation was observed between the magnetic field and qubit frequency shift or dephasing parameters. Therefore, our analysis focused solely on \( T_1 \) and its fluctuations.

\subsection{Impact of Trapped Magnetic Flux}

\begin{figure}[t]
\centering\includegraphics[width=\columnwidth]{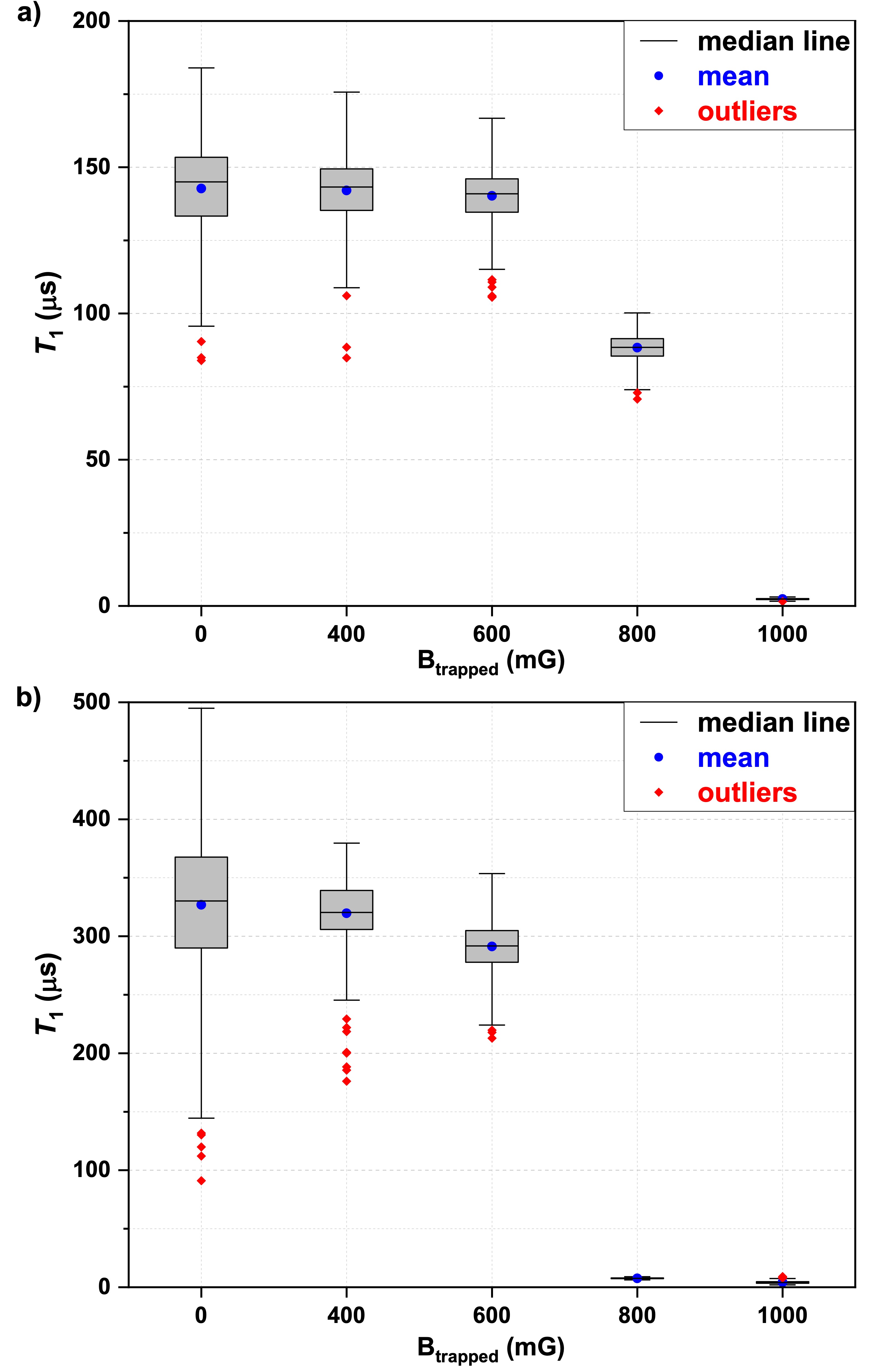}
\caption{Box plots of \( T_1 \) under varying magnetic fields (\( B_\text{trapped} \)) applied during cooldown, illustrating the impact of trapped flux on qubit performance:
a) \( T_1 \) distribution for \( q_1 \), and  
b) \( T_1 \) distribution for \( q_2 \).  
The interquartile range and mean values shown in each plot highlight a clear degradation of \( T_1 \) for \( B_\text{trapped} > 600\,\mathrm{mG} \), and a reduction in temporal fluctuations with increased trapped magnetic field.
}
\label{FIG:boxplot q1 and q2}
\end{figure}

To assess the influence of trapped magnetic flux on qubit relaxation times and their stability, we measured \( T_1 \) for qubits \( q_1 \) and \( q_2 \) after cooldowns performed under various static magnetic fields \( B_\text{trapped} \) (0--1000\,mG). The results are summarized in TABLE~\ref{tab:q1 and q2} and visualized in FIG.~\ref{FIG:boxplot q1 and q2} with the box plot.

At zero trapped field, \( q_1 \) exhibited a mean \( T_1 \) of 142.7\,µs and MAD of 10.1\,µs, while \( q_2 \) showed a significantly longer \( T_1 \) of 326.9\,µs with MAD of 38.0\,µs. With moderate trapped fields (400--600\,mG), both qubits displayed relatively stable \( T_1 \) values, with slight decreases in the mean and notable improvements in temporal stability. At 600\,mG, \( q_1 \) maintained a mean \( T_1 \) of 140.2\,µs with a reduced MAD of 5.2\,µs; \( q_2 \) showed a mean \( T_1 \) of 291.3\,µs and MAD of 13.7\,µs.

However, at higher field strengths, performance degraded sharply. At 800\,mG, \( q_1 \)'s \( T_1 \) dropped to 88.3\,µs (MAD = 3.0\,µs), while \( q_2 \) showed a dramatic suppression to 7.6\,µs (MAD = 0.3\,µs). At 1000\,mG, relaxation times were almost eliminated, with \( q_1 \) and \( q_2 \) showing mean values of \( T_1 \) of 2.4\,µs and 4.2\,µs, respectively.

These results reveal that both qubits tolerate flux trapping up to about 600\,mG with modest \( T_1 \) degradation and improved stability. Beyond this threshold, a sharp transition into a dissipation-dominated regime is observed. This suggests a critical trapping field between 600--800\,mG beyond which coherence is severely compromised.

\begin{table}[h]
\centering
\caption{Summary of relaxation times for \( q_1 \) and \( q_2 \) under different magnetic fields \( B_\text{trapped} \) applied during cooldown from 15\,K to 3\,K. Mean (M) and mean absolute deviation (MAD) are shown.}
\label{tab:q1 and q2}
\begin{tabular}{cccccc}
\toprule
\textbf{\( B_\text{trapped} \) (mG)} & \multicolumn{2}{c}{\textbf{\( q_1 \)}} & \multicolumn{2}{c}{\textbf{\( q_2 \)}} \\
 & \( M \) (µs) & MAD (µs) & \( M \) (µs) & MAD (µs) \\
\midrule
0    & 142.7 & 10.1 & 326.9 & 38.0 \\
400  & 142.1 & 7.2  & 319.6 & 16.7 \\
600  & 140.2 & 5.2  & 291.3 & 13.7 \\
800  & 88.3  & 3.0  & 7.6   & 0.3 \\
1000 & 2.4   & 0.2  & 4.2   & 0.7 \\
\bottomrule
\end{tabular}
\end{table}

\subsection{Temperature Dependence in the Flux-Trapped Regime}

\begin{figure}[h]
\centering\includegraphics[width=\columnwidth]{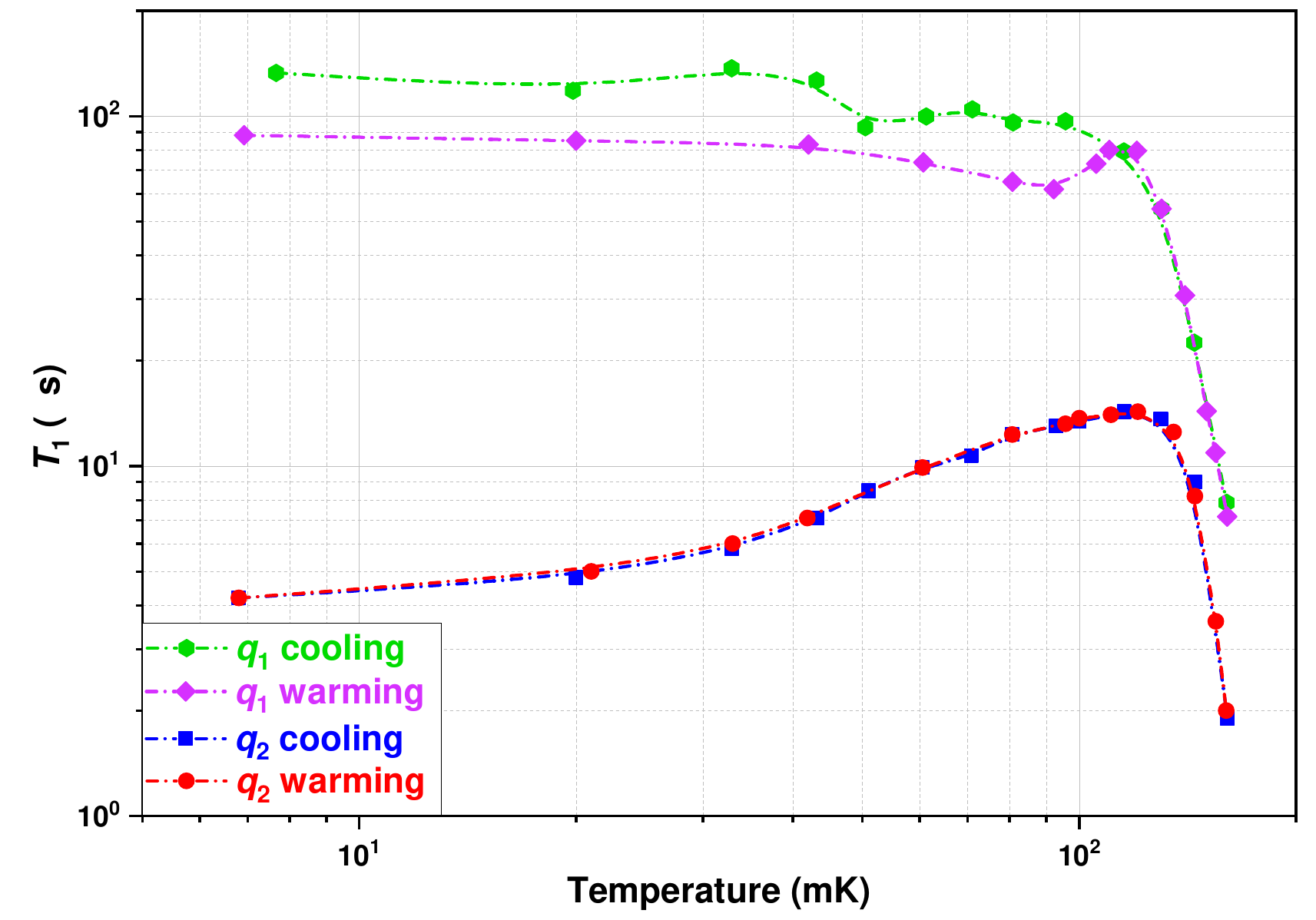}
\caption{Temperature dependence of energy relaxation time for $q_1$ and $q_2$ after trapping magnetic flux with an 800 mG field during cooldown. $T_1$ was measured as the temperature was swept from 8 mK to 200 mK and back down to 8 mK.}
\label{tsweep}
\end{figure}

Temperature sweeps were performed after trapping magnetic flux using an 800\,mG field for \( q_1 \) and \( q_2 \). The results are presented in FIG.~\ref{tsweep}.

Above approximately 125\,mK, both qubits exhibited a sharp drop in \( T_1 \). Below this threshold, \( q_1 \) showed relatively stable relaxation times during both heating and cooling, indicating resilience to moderate thermal perturbations in the flux-trapped regime.

Below 125\,mK, \( q_2 \) displayed an anomalous trend: \( T_1 \) continued to decline during both warming and cooling, suggesting a persistent dissipation mechanism which may be attributed to trapped flux-induced losses or strong coupling to TLS.

\subsection{Influence of Actively Applied Magnetic Fields}

\begin{figure}[h]
\centering\includegraphics[width=\columnwidth]{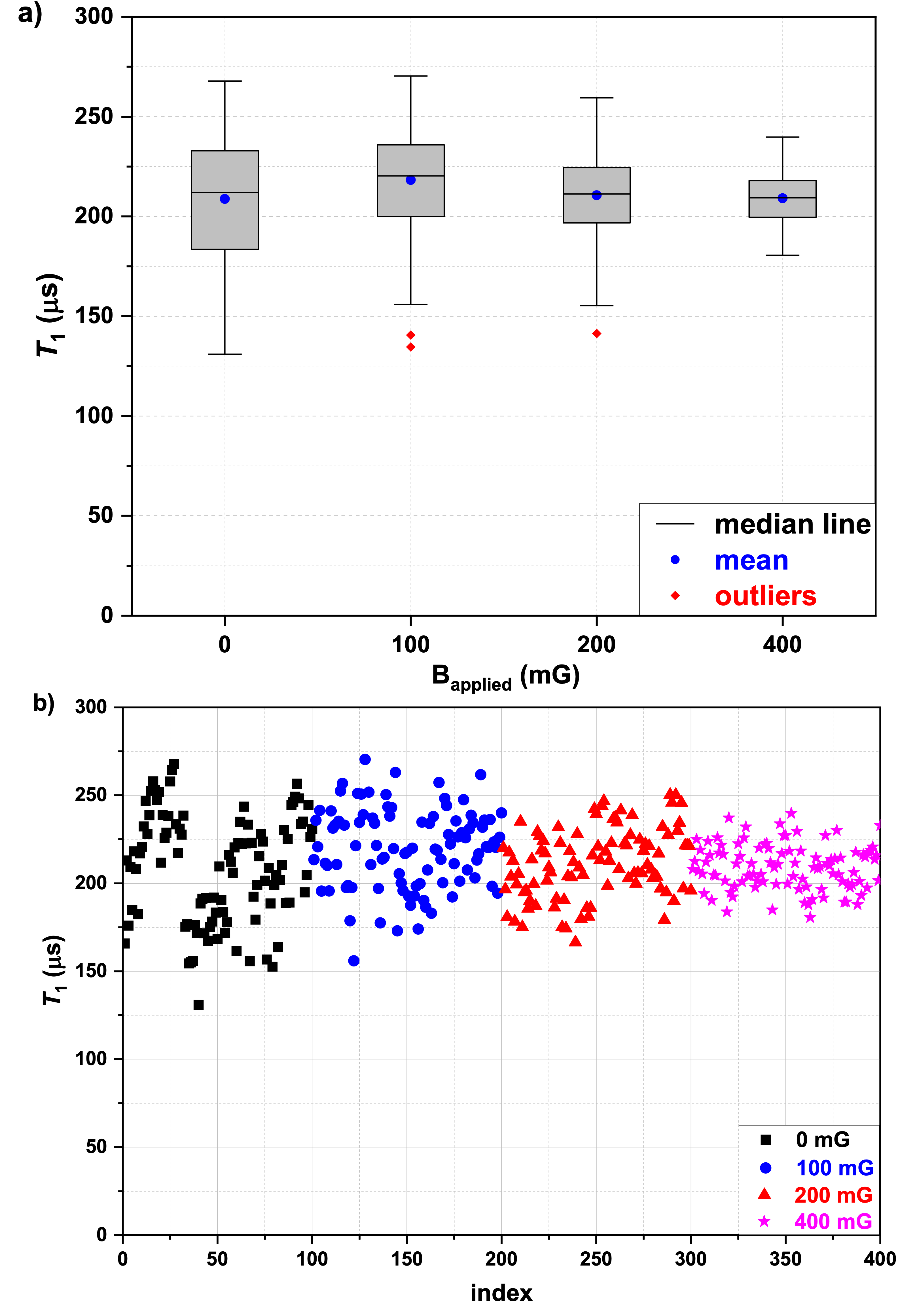}
\caption{a) Box plot of $T_1$ values for $q_3$ under actively applied magnetic fields (\( B_\text{applied} \)) of 0, 100, 200, and 400 mG. The reduction in spread indicates enhanced temporal stability with increasing field strength. b) Scatter plot of individual $T_1$ measurements over time for each field condition, showing the evolution of temporal fluctuations during the 15-hour measurement windows. A progressive suppression of noise is observed with increasing \( B_\text{applied} \).}
\label{q3}
\end{figure}

We also studied the effect of actively applied static magnetic fields at base temperature on \( q_3 \), applying perpendicular fields (along the \( z \)-axis) of 0, 100, 200, and 400\,mG. Unlike the flux-trapping experiments, the device was cooled to base temperature in zero magnetic field, and the magnetic field was applied only during \( T_1 \) measurements at 8\, mK.
The box plot and scatter plot of the measurements are illustrated in FIG.~\ref{q3}.

At zero field, \( q_3 \) showed a mean \( T_1 \) of 208.7\,µs with MAD of 23.2\,µs. Applying a 100\,mG field slightly improved \( T_1 \) to 218.2\,µs and reduced MAD to 17.4\,µs. At 200\,mG, the mean \( T_1 \) remained stable at 209.1\,µs, with a further reduction in MAD to 13.7\,µs. Increasing the field to 400\,mG decreased \( T_1 \) slightly to 205.4\,µs, while continuing to suppress fluctuations (MAD = 10.7\,µs).

These results suggest that moderate magnetic fields applied during operation may stabilize qubit behavior by suppressing certain noise mechanisms. Further increases in field strength were not explored due to heating limitations in the Helmholtz coil system.

\section{Discussion}

The impact of trapped magnetic flux on qubit coherence exhibits a nuanced, threshold-like behavior. While earlier studies on niobium SRF cavities show gradual degradation with increased magnetic field during cooldown (due to the increasing fluxoid losses trapped in the niobium surface) ~\cite{bafia2025quantifying}, our experimental results for transmon qubits reveal a sharp transition. For trapped fields up to approximately \(600\,\mathrm{mG}\), both qubits (\(q_1\) and \(q_2\)) maintain high \(T_1\) values with reduced temporal fluctuations, suggesting that flux-induced dissipation remains negligible in this regime. However, beyond a critical range, between \(600\,\mathrm{mG}\) and \(800\,\mathrm{mG}\), we observe a sharp decline in \(T_1\), particularly pronounced in \(q_2\). This abrupt degradation points to the onset of vortex-related losses, which cannot be explained only by flux losses in the capacitor pads, and could hint to flux abruptly entering into the junction area. Under a trapped field of \(1000\,\mathrm{mG}\), relaxation times are significantly reduced, with \(T_1\) values dropping to \(2.4\,\mu\mathrm{s}\) for \(q_1\) and \(4.2\,\mu\mathrm{s}\) for \(q_2\), underscoring a substantial loss in coherence attributable to flux trapping and vortex dissipation mechanisms.

Earlier studies reported flux thresholds of ~300\,mG for aluminum and ~450\,mG for rhenium~\cite{song2009microwave}; however, the threshold for Nb capped with Ta structures is strongly dependent on material properties and geometry~\cite{PhysRevB.91.180505}. Our data suggest a threshold between 600–800\,mG, above which flux-induced losses dominate.

Previous works suggest vortices can act as sinks for non-equilibrium QPs that tunnel through the JJ ~\cite{wang2014measurement,schneider2019transmon,PhysRevLett.106.077002}. This vortex-induced QP trapping may mitigate the loss of coherence, depending on the vortex distribution and interaction with other dissipation channels. Non-equilibrium QPs generated by warming the chip after trapping flux with 800\,mG showed interesting results. Above 125\,mK, both \( q_1 \) and \( q_2 \) show a sharp decline in \( T_1 \), consistent with thermally activated QP generation in aluminum-based JJs~\cite{PhysRevLett.106.077002,PhysRevLett.107.240501}. Below 125\,mK, \( q_1 \)'s \( T_1 \) stabilizes, indicating dominant but relatively temperature-insensitive loss mechanisms. In contrast, \( q_2 \)'s \( T_1 \) continues to decline with decreasing temperature, which may arise due to two potential sources: TLS-driven loss due to surface oxide~\cite{abdisatarov2024direct,Reagor2016SuperconductingCF} or thermal activation of pinning flux centers, recently identified in niobium SRF cavities \cite{bafia2025quantifying}. The observed difference in behaviors between the two qubits likely arises from \( q_2 \)'s larger capacitor pads and wider spacing (180\,\textmu m vs.\ 20\,\textmu m).

Importantly, we find that trapped and applied magnetic fields suppress \( T_1 \) fluctuations for the qubits studied here (see FIG.~\ref{FIG:boxplot q1 and q2} and ~\ref{q3}). For \( q_3 \), applying static magnetic field up to 400\,mG during \( T_1 \) measurements had minimal effect on mean \( T_1 \) but reduced temporal fluctuation amplitudes. This stabilization likely arises from paramagnetic impurity polarization (O\(_2\), NbO, TaNb), trapping QPs and certain part of TLS saturation.

To probe temporal noise, we applied Allan deviation analysis—a time-domain metrology technique effective in identifying stochastic processes~\cite{allan1966statistics}. For \( q_2 \), under 0\,mG, 400\,mG, and 600\,mG, we used:
\[
\sigma(\tau) = \left(\frac{n_0}{2}\right)^{1/2} \tau^{-1/2} + \left(2 \ln 2 \cdot n_1\right)^{1/2} + \left(\frac{4 \pi^2}{6} n_2\right)^{1/2} \tau^{1/2},
\]
where \( n_0, n_1, n_2 \) denote white, flicker, and random walk noise amplitudes ~\cite{ieee1998gyro,yang2023locating}.

\begin{table}[ht]
\centering
\caption{Fitted noise amplitudes from Allan deviation analysis of \( q_2 \) under varying trapped flux.}
\label{al}
\begin{tabular}{c|c|c|c}
\hline
\( B_{\mathrm{trapped}} \) (mG) & \( n_0 \) (\( \times 10^5 \)) & \( n_1 \) (\( \times 10^{-2} \)) & \( n_2 \) (\( \times 10^{-3} \)) \\
\hline
0   & 8.47 & 2.61 & 3.01 \\
400 & 1.01 & 2.60 & 2.95 \\
600 & 0.95 & 0.42 & 0.49 \\
\hline
\end{tabular}
\end{table}

\begin{figure}[h]
\centering\includegraphics[width=\columnwidth]{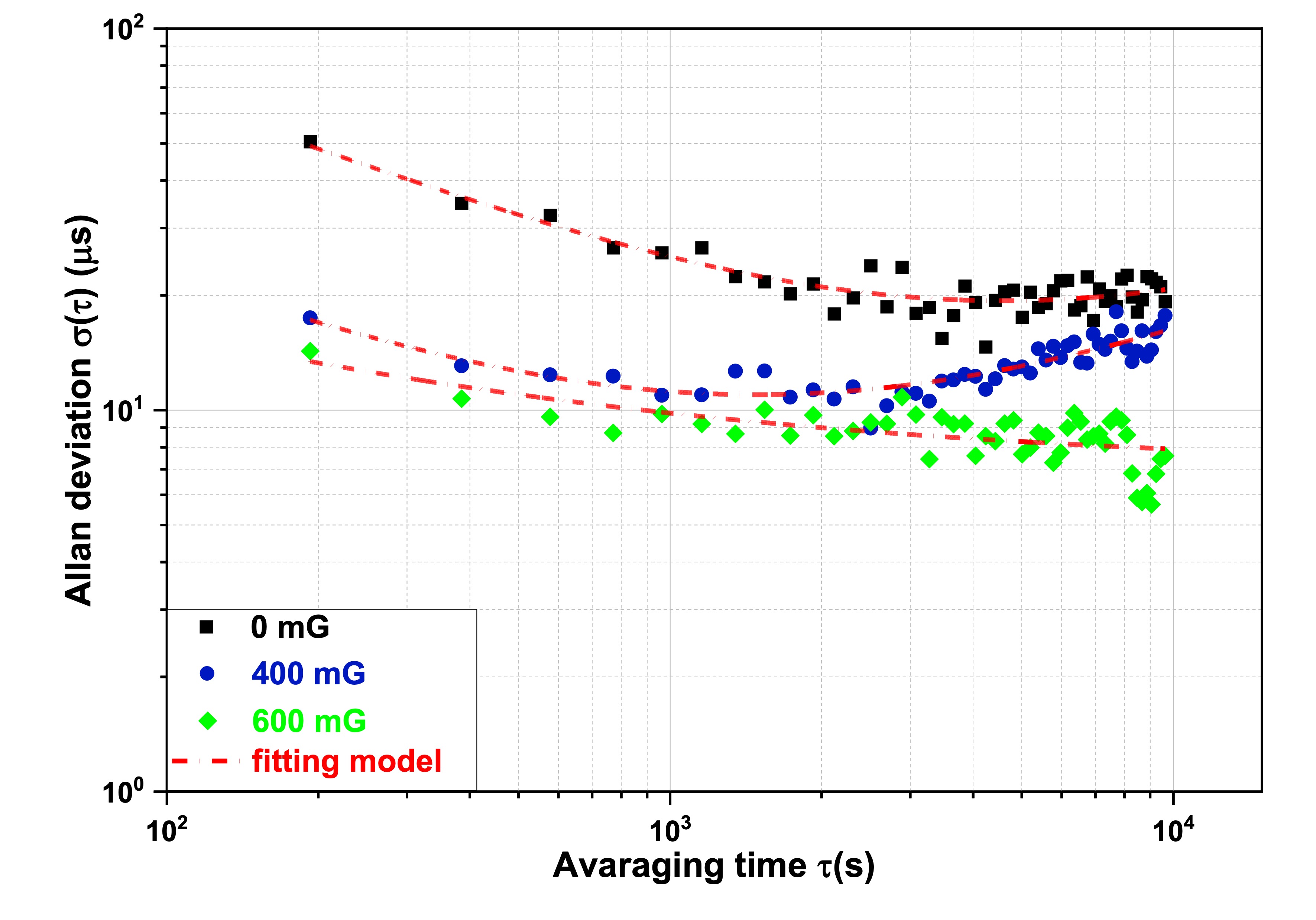}
\caption{Allan deviation analysis and noise model fitting of qubit $q_2$ under varying trapped magnetic field conditions. Black squares represent data at $0\,\mathrm{mG}$, blue circles at $400\,\mathrm{mG}$, and green diamonds at $600\,\mathrm{mG}$. The red dashed line indicates the total fitted noise model.}
\label{allan}
\end{figure}

The Allan deviation fitting results, illustrated in FIG.~\ref{allan} and summarized in TABLE~\ref{al}, reveal a consistent and substantial suppression of noise amplitudes with increasing trapped magnetic flux. At 0\,mG, the dominant white noise amplitude \( n_0 \) is high (\( 8.47 \times 10^5 \)), indicative of significant short-term instability. This elevated noise level is likely driven by rapid fluctuations from high-frequency TLS~\cite{klimov2018fluctuations, PRXQuantum.3.040332, carroll2022dynamics}, non-equilibrium QPs tunneling through the JJ~\cite{catelani2012decoherence,
aumentado2023quasiparticle, wang2014measurement}, and magnetic noise stemming from paramagnetic impurities~\cite{lee2014identification, sendelbach2008magnetism, cava1991electrical, pritchard2024suppressed}.

With the introduction of trapped magnetic flux at 400\,mG and 600\,mG, \( n_0 \) drops by nearly an order of magnitude, providing strong evidence of effective suppression of high-frequency noise. This reduction can be attributed to three key mechanisms: (i) partial polarization of paramagnetic impurities (e.g., O\(_2\), NbO, or TaNb alloys), which diminishes their high-frequency magnetic susceptibility; (ii) quasiparticle trapping in vortex cores, thereby lowering the population of mobile QPs near the junction; and (iii) saturation of specific high-frequency TLS loss channels.

The flicker noise component \( n_1 \), typically associated with low-frequency magnetic fluctuations and TLS dynamics, remains relatively stable at 400\,mG but shows a notable decrease at 600\,mG. This trend implies that higher levels of trapped flux begin to influence low-frequency noise sources, possibly through enhanced TLS saturation or further polarization of paramagnetic species. Likewise, the random walk noise component \( n_2 \), which reflects long-term drift phenomena, decreases with increasing trapped flux, suggesting improved long-term qubit stability and reduced sensitivity to slow environmental fluctuations.

These results support a framework where controlled flux trapping or applied magnetic fields mitigate QP-related decoherence and quantum noise. By engineering the magnetic environment, we significantly reduce \( T_1 \) fluctuations through the simultaneous suppression of non -equilibrium QPs, TLSs, and magnetic defects.

\vspace{20pt} 
\section{Conclusion}
 In summary, we demonstrate that transmon qubits based on Nb/Ta capacitor structures not only tolerate moderate trapped or applied magnetic fields, but can also benefit from them. Trapped fields below a critical threshold (\(\sim600\,\mathrm{mG}\)) and applied fields below (\(\sim400\,\mathrm{mG}\)) suppress temporal fluctuations in energy relaxation time without degrading its mean value, an effect we attribute to the polarization of paramagnetic impurities, the trapping of non-equilibrium QPs, and the saturation of certain TLS losses. Beyond this threshold, coherence rapidly deteriorates, revealing a sharp boundary between beneficial and deleterious flux regimes. This threshold can be modified and optimized in future experiments, to allow for further improvements in coherence fluctuations reduction and control while maintaining high average coherence values. 

These findings introduce a new paradigm: magnetic fields, when precisely controlled, are not merely a background disturbance to be shielded against, but can serve as a tool for engineering more stable qubit environments. This work opens a new frontier in decoherence mitigation, with immediate implications for the design, calibration, and scalability of next-generation superconducting quantum processors.

\vspace{15pt} 
\acknowledgments
The work was supported by the U.S. Department of Energy, Office of Science, National Quantum Information Science Research Centers, Superconducting Quantum Materials and Systems (SQMS) Center under the contract No. DE-AC02-07CH11359. The authors thank the Director and technical staff of the Pritzker Nanofabrication facility.

\vspace{15pt} 

\bibliographystyle{unsrt}
\bibliography{references}

\end{document}